\documentclass[english,aps,preprint,superscriptaddress,showpacs]{revtex4-2}
\usepackage[utf8]{inputenc}
\usepackage{color}
\usepackage{babel}
\usepackage{units}
\usepackage{bm}
\usepackage{amsmath}
\usepackage{amssymb}
\usepackage{graphicx}
\usepackage[unicode=true,
 bookmarks=true,bookmarksnumbered=false,bookmarksopen=true,bookmarksopenlevel=1,
 breaklinks=false,pdfborder={0 0 0},pdfborderstyle={},backref=false,colorlinks=true]
 {hyperref}
\hypersetup{pdftitle={Final-state interaction in the process e+e− → Λc Λ̄c},
 pdfauthor={S. G. Salnikov},
 citecolor=blue,urlcolor=blue,linkcolor=blue}

\makeatletter

\providecommand{\tabularnewline}{\\}

\usepackage{bm}

\makeatother

\begin{document}
\title{Final-state interaction in the process $e^{+}e^{-}\rightarrow\Lambda_{c}\bar{\Lambda}_{c}$}
\author{A.I. Milstein}
\email{A.I.Milstein@inp.nsk.su}

\author{S.G. Salnikov}
\email{S.G.Salnikov@inp.nsk.su}

\affiliation{Budker Institute of Nuclear Physics of SB RAS, 630090 Novosibirsk,
Russia}
\affiliation{Novosibirsk State University, 630090 Novosibirsk, Russia}
\date{\today}
\begin{abstract}
We show that the final-state interaction explains the nontrivial near-threshold
energy dependence of the cross section of the process $e^{+}e^{-}\rightarrow\Lambda_{c}\bar{\Lambda}_{c}$
observed by the Belle and BESIII collaborations. This energy dependence
is the result of the mixture of $S$-wave and $D$-wave components
of the $\Lambda_{c}\bar{\Lambda}_{c}$ wave function due to a tensor
interaction. The Coulomb potential is important only in the narrow
energy region about a few MeV above the threshold of the process.
It is shown that the widely used assumption that the impact of the
Coulomb interaction on the cross sections of hadron production is
reduced to the Sommerfeld-Gamow-Sakharov factor is not correct.
\end{abstract}
\maketitle

\section{Introduction}

During last decades a lot of processes with hadron production were
investigated in detail in the corresponding near-threshold energy
regions. In these energy regions a strong energy dependence of the
cross sections was observed. For instance, this phenomenon is manifested
in the processes %
\mbox{%
$e^{+}e^{-}\to p\bar{p}$%
}~%
\mbox{%
\citep{Aubert2006,Lees2013,Lees2013a,Akhmetshin2016,Akhmetshin2019,Ablikim2019}%
}, %
\mbox{%
$e^{+}e^{-}\to n\bar{n}$%
}~\citep{Achasov2014}, %
\mbox{%
$e^{+}e^{-}\to B\bar{B}$%
}~\citep{Achasov2018}, %
\mbox{%
$J/\psi(\psi')\to p\bar{p}\pi^{0}(\eta)$%
}~%
\mbox{%
\citep{Bai2001,Ablikim2009,Bai2003}%
}, %
\mbox{%
$J/\psi(\psi')\to p\bar{p}\omega(\gamma)$%
}~%
\mbox{%
\citep{Bai2003,Ablikim2008,Alexander2010,Ablikim2012,Ablikim2013b}%
}, and %
\mbox{%
$e^{+}e^{-}\to\phi\Lambda\bar{\Lambda}$%
}~\citep{Ablikim2021c}. Note that there is no conventional view on
the origin of such strong energy dependence. One of the most natural
explanations is the effect of the  final-state interaction of produced
hadrons. Indeed, a small relative velocity of hadrons in the near-threshold
energy region results in strong effect due to large effective time
of interaction. In a set of publications~%
\mbox{%
\citep{dmitriev2007final,dmitriev2014isoscalar,Dmitriev2016,Milstein2018,Milstein2021,Haidenbauer2014,Kang2015,Dmitriev2016a,Milstein2017,Milstein2021c}%
} it was shown that the account for the final-state interaction allows
one to explain the available experimental data with good accuracy.
At present, it is impossible to describe interaction of hadrons at
small relative velocities using QCD. As a result, it is necessary
to employ various phenomenological models. Comparison of the theoretical
predictions with the available experimental data allows one to fix
the parameters of the models.

Much attention of researchers was attracted to the process $e^{+}e^{-}\to\Lambda_{c}\bar{\Lambda}_{c}$.
The corresponding cross section demonstrates a very nontrivial energy
dependence in the vicinity of the threshold. The Belle Collaboration
observed a peak in the cross section at the energy of $\Lambda_{c}\bar{\Lambda}_{c}$
pair about $\unit[80]{MeV}$ above the threshold~\citep{Pakhlova2008}.
Later the data obtained by the %
\mbox{%
BESIII%
} Collaboration demonstrated a plateau in the cross section in the
energy region from $\unit[1.5]{MeV}$ to $\unit[30]{MeV}$ above the
threshold~\citep{Ablikim2018b}. At first glance these two sets of
data seems to be inconsistent with each other, and it is not clear
if all these data can be explained by the final-state interaction
of $\Lambda_{c}$~baryons~\citep{Dai2017a}. Recently there was
an attempt to describe the behavior of the cross section of this process
using the modified Sommerfeld-Gamow-Sakharov factor~\citep{Amoroso2021}.
However, this factor alone can not describe the peak in the cross
section at the energy about $\unit[80]{MeV}$. It is necessary to
emphasize that the effective model of $\Lambda_{c}\bar{\Lambda}_{c}$
interaction should describe not only the energy dependence of the
cross section but also the energy dependence of the ratio $\left|G_{E}/G_{M}\right|$
of electromagnetic form factors of $\Lambda_{c}$~baryon. This ratio
was also measured by the BESIII Collaboration~\citep{Ablikim2018b}.
In the present work we propose a simple model of $\Lambda_{c}\bar{\Lambda}_{c}$
interaction which reproduces all features of the cross section of
the process $e^{+}e^{-}\to\Lambda_{c}\bar{\Lambda}_{c}$ together
with the energy dependence of the ratio $\left|G_{E}/G_{M}\right|$.

\section{Theoretical approach}

The method to account for the effect of baryon-antibaryon final-state
interaction was developed in Refs.~%
\mbox{%
\citep{dmitriev2007final,dmitriev2014isoscalar,Dmitriev2016,Milstein2018}%
} for the case of nucleon-antinucleon system. This method is based
on the assumption that the process of production of non-relativistic
hadrons can be separated into two stages. At the first stage, virtual
hadrons are produced at small distances, and the amplitude of their
production is a smooth function of the energy of the system. At the
second stage, the interaction takes place at large distances where
the hadrons become real, but not virtual. Therefore, any sharp behavior
of the cross section of the process is the result of the interaction
of hadrons at large distances. This interaction can be described by
some effective optical potentials. The imaginary part of the optical
potentials takes into account the annihilation of hadrons into mesons.
In the case of nucleon-antinucleon pair production~%
\mbox{%
\citep{dmitriev2007final,dmitriev2014isoscalar,Dmitriev2016,Milstein2018}%
}, it is necessary to account for the components of the wave function
with the isospins $I=0$ and $I=1$. A~mixture of these components
arises, firstly, due to the electromagnetic interaction, and secondly,
due to the difference of proton and neutron masses. Consideration
of the process $e^{+}e^{-}\to\Lambda_{c}\bar{\Lambda}_{c}$ is essentially
simpler than that of the processes of nucleon-antinucleon pair production.
The effective potential of $\Lambda_{c}\bar{\Lambda}_{c}$ interaction
is real because it is not necessary to account for the $\Lambda_{c}\bar{\Lambda}_{c}$
annihilation into mesons. Besides, the isospin of $\Lambda_{c}\bar{\Lambda}_{c}$
pair is zero.

The process of $e^{+}e^{-}$ annihilation into $\Lambda_{c}\bar{\Lambda}_{c}$
pair goes through a virtual photon. Hence, the quantum numbers of
the pair are $J^{PC}=1^{--}$, so that the angular momentum of a pair
is $l=0,2$ and the total spin is $S=1$. The $S$-wave and $D$-wave
components of the wave function are mixed by the tensor forces. The
effective potential of $\Lambda_{c}\bar{\Lambda}_{c}$ interaction
contains several parts and has the form ($\hbar=c=1$)
\begin{equation}
\mathcal{V}(r)=-\frac{\alpha}{r}+V_{S}(r)\,\delta_{l0}+\left(\frac{6}{Mr^{2}}+V_{D}(r)\right)\delta_{l2}+V_{T}(r)\,S_{12}\,.
\end{equation}
Here $\alpha$~is the fine-structure constant, $V_{S}$, $V_{D}$,
and $V_{T}$~are the $S$-wave, the $D$-wave, and the tensor contributions
to the potential, respectively, $S_{12}=6\left(\bm{S}\cdot\bm{n}\right)^{2}-4$
is the tensor operator, $\bm{S}$~is the spin operator of the $\Lambda_{c}\bar{\Lambda}_{c}$
pair, and~%
\mbox{%
$\bm{n}=\bm{r}/r$%
}. The corresponding coupled-channels radial Schr\"{o}dinger equation
can be written in the form
\begin{equation}
\left[\frac{p_{r}^{2}}{M}+\mathcal{V}(r)-E\right]\Psi(r)=0\,,\label{eq:schrodinger}
\end{equation}
where $M$~is the mass of $\Lambda_{c}$ baryon, $E$~is the energy
of the pair, counted from the threshold, and $\left(-p_{r}^{2}\right)$~is
the radial part of the Laplace operator. The wave function $\Psi(r)$
of the Schr\"{o}dinger equation~(\ref{eq:schrodinger}) has two
components, namely, $\Psi^{T}(r)=\left(u(r),\,w(r)\right)$, where
the first component corresponds to the $S$-wave and the second one
to the $D$-wave. In this basis, the potential $\mathcal{V}(r)$ can
be written in a matrix form
\begin{equation}
\mathcal{V}(r)=\begin{pmatrix}-\frac{\alpha}{r}+V_{S} & -2\sqrt{2}V_{T}\\
-2\sqrt{2}V_{T}\quad & -\frac{\alpha}{r}+\frac{6}{Mr^{2}}+V_{D}-2V_{T}
\end{pmatrix}.
\end{equation}

Two regular independent solutions of the Schr\"{o}dinger equation~(\ref{eq:schrodinger})
have the following asymptotics at $r\to\infty$ (see Refs.~%
\mbox{%
\citep{dmitriev2007final,dmitriev2014isoscalar,Dmitriev2016,Milstein2018}%
}) 
\begin{align}
 & \Psi_{1}^{T}(r)=\frac{1}{2i}\left(S_{11}\chi_{0}^{+}-\chi_{0}^{-},\,S_{12}\chi_{2}^{+}\right),\nonumber \\
 & \Psi_{2}^{T}(r)=\frac{1}{2i}\left(S_{21}\chi_{0}^{+},\,S_{22}\chi_{2}^{+}-\chi_{2}^{-}\right),\nonumber \\
 & \chi_{l}^{\pm}=\frac{1}{kr}\exp\left[\pm i\left(kr-l\pi/2+\eta\ln\left(2kr\right)+\sigma_{l}\right)\right],\nonumber \\
 & \sigma_{l}=\frac{i}{2}\ln\frac{\Gamma\left(1+l+i\eta\right)}{\Gamma\left(1+l-i\eta\right)}\,,\qquad\eta=\frac{M\alpha}{2k}\,,\qquad k=\sqrt{ME}\,,
\end{align}
where $\Gamma(x)$~is the Euler gamma function and $S_{ij}$~are
some functions of energy. In the non-relativistic approximation, the
electric $G_{E}$ and the magnetic $G_{M}$ form factors of $\Lambda_{c}$
baryon in the time-like region are expressed in terms of $u_{1}(0)$
and $u_{2}(0)$ which are the $S$-wave components of two independent
solutions at $r=0$:
\begin{align}
 & G_{E}=\mathcal{G}\left(u_{1}(0)-\sqrt{2}\,u_{2}(0)\right),\nonumber \\
 & G_{M}=\mathcal{G}\left(u_{1}(0)+\frac{1}{\sqrt{2}}u_{2}(0)\right).
\end{align}
Here $\mathcal{G}$~is the energy independent amplitude of $\Lambda_{c}\bar{\Lambda}_{c}$
pair production at small distances. Note that $u_{2}(0)$ is nonzero
only due to account for the tensor forces and $D$-wave component
of the wave function. The energy dependence of the ratio $G_{E}/G_{M}$
is determined by the energy dependence of the ratio $f=u_{2}(0)/u_{1}(0)$:
\begin{equation}
\frac{G_{E}}{G_{M}}=\frac{1-\sqrt{2}\,f}{1+\frac{1}{\sqrt{2}}f}\,.
\end{equation}
The integrated cross section of $\Lambda_{c}\bar{\Lambda}_{c}$ pair
production has the form
\begin{equation}
\sigma=\frac{\pi k\alpha^{2}}{2M^{3}}\left|\mathcal{G}\right|^{2}\left(\left|u_{1}(0)\right|^{2}+\left|u_{2}(0)\right|^{2}\right).
\end{equation}
Therefore, near the threshold both the cross section and the ratio
of electromagnetic form factors depend on the energy via the functions
$u_{1}(0)$ and $u_{2}(0)$. In the present paper we calculate numerically
these functions using some effective potential. The parameters of
this potential are fixed by fitting the available experimental data.

\section{Results and discussion}

The exact potential of $\Lambda_{c}\bar{\Lambda}_{c}$ interaction
is unknown, so a phenomenological potential model has to be proposed.
Our previous works~%
\mbox{%
\citep{Milstein2021,Milstein2021c}%
} devoted to the final-state interaction in various hadronic systems
showed that the enhancement of the cross section of hadronic pair
production is usually associated with existence of a near-threshold
resonant state. The shape of the invariant-mass spectrum of hadronic
pair production is determined mainly by the parameters of this resonance,
and the specific parameterization of the potential is not so important.
Therefore, we consider the $S$-wave, $D$-wave, and tensor parts
of the potential as rectangular potential wells:
\begin{equation}
V_{n}(r)=U_{n}\,\theta(a_{n}-r)\,,\qquad n=S,\,D,\,T\,,
\end{equation}
where $\theta(x)$~is the Heaviside function, $U_{n}$ and $a_{n}$~are
some fitting parameters. In addition, for convenience of numerical
calculations the tensor potential is regularized at small distances
by the factor
\begin{equation}
F(r)=\frac{(br)^{2}}{1+(br)^{2}}
\end{equation}
with $b=\unit[10]{fm^{-1}}$. In fact, the results are almost independent
of the specific value of the parameter~$b$. The parameters of the
potential, as well as the coefficient~$\mathcal{G}$, are determined
by fitting the experimental data and minimizing~$\chi^{2}$. The
experimental data includes measurements of the cross section of the
process %
\mbox{%
$e^{+}e^{-}\to\Lambda_{c}\bar{\Lambda}_{c}$%
} collected by the Belle~\citep{Pakhlova2008} and BESIII~\citep{Ablikim2018b}
collaborations, as well as two measurements for the ratio of electric
and magnetic form factors of $\Lambda_{c}$ baryon obtained by BESIII~\citep{Ablikim2018b}.

\begin{figure}
\includegraphics[totalheight=5.4cm]{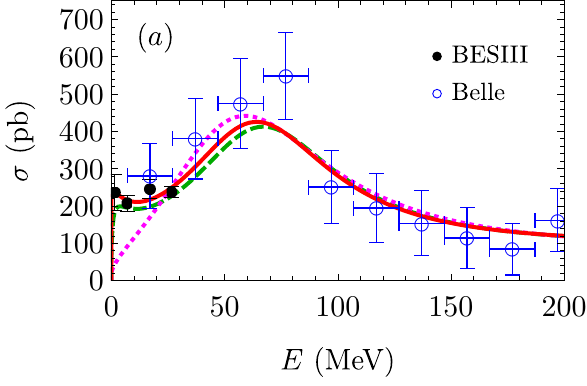}\hfill{}\includegraphics[totalheight=5.4cm]{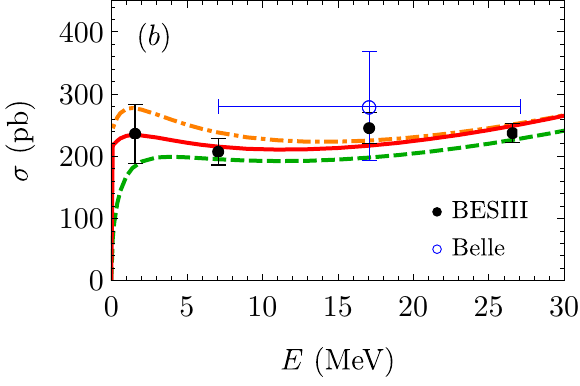}
\begin{centering}
\includegraphics[totalheight=5.4cm]{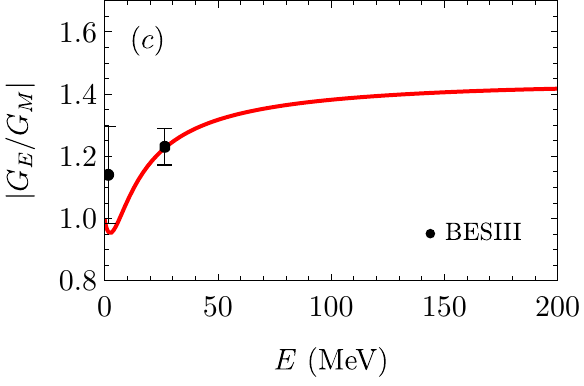}
\par\end{centering}
\caption{The cross section of the process $e^{+}e^{-}\to\Lambda_{c}\bar{\Lambda}_{c}$
in the energy region from the threshold to $\unit[200]{MeV}$, Fig.~(a),
and in the narrow energy region, Fig.~(b). The ratio of electric
and magnetic form factors of $\Lambda_{c}$ baryon is shown in Fig.~(c).
The solid curves correspond to our predictions obtained with all contributions
taken into account. The dashed curves are obtained without accounting
for the Coulomb potential. The dotted curve corresponds to the prediction
for zero value of the tensor potential. The dash-dotted curve shows
the result for zero Coulomb potential multiplied by the Sommerfeld-Gamow-Sakharov
factor. The experimental data are from Refs.~%
\mbox{%
\citep{Pakhlova2008,Ablikim2018b}%
}.}\label{fig:sigGeGm}
\end{figure}

The parameters of the potential corresponding to the best fit are
listed in Table~\ref{tab:params}. Note that the radii of the $D$-wave
and tensor parts of the potential turned out to be close to each other,
and we set them to be equal. For the parameters of the potential obtained
within our approach, the value of $\chi^{2}$ is $7.5$, so that $\chi^{2}/N_{df}=0.75$,
where $N_{df}$~is the number of degrees of freedom. The results
of this fit are shown in Fig.~\ref{fig:sigGeGm} by the solid curves.

\tabcolsep=2ex

\begin{table}
\begin{centering}
\begin{tabular}{|l|c|c|c|}
\hline 
 & $V_{S}$ & $V_{D}$ & $V_{T}$\tabularnewline
\hline 
$\unit[U]{(MeV)}$ & $-447_{-4.1}^{+5.1}$ & $363_{-33}^{+42}$ & $22.1_{-1.2}^{+1.1}$\tabularnewline
\hline 
$\unit[a]{(fm)}$ & $1.425_{-0.007}^{+0.006}$ & $2.66_{-0.09}^{+0.1}$ & $2.66_{-0.09}^{+0.1}$\tabularnewline
\hline 
\end{tabular}
\par\end{centering}
\caption{The parameters of the potential of $\Lambda_{c}\bar{\Lambda}_{c}$
interaction.}\label{tab:params}
\end{table}

Let us discuss the effect of various contributions to the potential
on the shape of the cross section and the ratio $\left|G_{E}/G_{M}\right|$.
If we set the tensor potential to be zero, then $\left|G_{E}/G_{M}\right|$
will be unity and the plateau in the cross section at energy below
$\unit[30]{MeV}$ will disappear. However, the peak at energy around
$\unit[70]{MeV}$ and its shape are well reproduced. The corresponding
results are shown in Fig.~\ref{fig:sigGeGm}a by the dotted curve.
Only all nonzero values of $V_{S}$, $V_{D}$, and $V_{T}$ allow
us to reproduce all set of experimental data.

It is interesting to investigate the effect of the Coulomb potential.
The cross section calculated without the Coulomb potential is shown
in Fig.~\ref{fig:sigGeGm} by the dashed curves. It is seen that
the Coulomb potential is important only in the energy region very
close to the threshold. For energies of $\Lambda_{c}\bar{\Lambda}_{c}$
pair above a few MeV the impact of the Coulomb interaction is not
very important. It is generally accepted that the cross section calculated
with the Coulomb interaction taken into account can be represented
as the cross section calculated without the Coulomb potential multiplied
by the so-called Sommerfeld-Gamow-Sakharov factor~$C$,
\begin{equation}
C=\frac{2\pi\eta}{1-e^{-2\pi\eta}}\,,\qquad\eta=\frac{M\alpha}{2k}\,.
\end{equation}
The cross section calculated using this approach is shown in Fig.~\ref{fig:sigGeGm}b
by the dash-dotted curve. Obviously, this result is quite different
from the result of the exact calculations (the solid curve). Despite
the fact that the Sommerfeld-Gamow-Sakharov factor provides a nonzero
cross section at the threshold, the factorization of the cross section
does not work well enough. The same conclusion was previously made
in Ref.~\citep{Milstein2018} when analyzing the cross section of
the process $e^{+}e^{-}\to p\bar{p}$. Therefore, absence of factorization
is not a specific feature of some process, but a general statement.

\section{Conclusion}

It is shown that near the threshold the nontrivial energy dependence
of the cross section of the process $e^{+}e^{-}\to\Lambda_{c}\bar{\Lambda}_{c}$
and the ratio $\left|G_{E}/G_{M}\right|$ can be well described by
the final-state interaction. We used a simple potential of $\Lambda_{c}\bar{\Lambda}_{c}$
interaction containing $S$-wave, $D$-wave and the tensor parts.
Each part of the potential was parameterized by a rectangular potential
well. A peak in the spectrum of the process corresponds to the near-threshold
resonant state of $\Lambda_{c}\bar{\Lambda}_{c}$ pair. The plateau
in the energy region below $\unit[30]{MeV}$ is due to the tensor
and $D$-wave parts of the potential. These parts of the potential
are responsible also for deviation of the ratio $\left|G_{E}/G_{M}\right|$
from unity.

\bibliographystyle{/home/sergey/Documents/BINP/BibTeX/apsrev4-1}

\end{document}